\documentclass[prd,onecolumn,superscriptaddress,preprintnumbers,showpacs,nofootinbib,notitlepage,floatfix]{revtex4-1}
\usepackage{hyperref}
\usepackage{graphicx}
\usepackage{subfigure}
\usepackage{amsmath}
\usepackage{amssymb}
\usepackage{slashed}
\usepackage{amsfonts}
\usepackage{xcolor}
\usepackage{multirow}
\usepackage{array}
\usepackage{mwe}
\usepackage{mathrsfs}

\begin{document}


\title{On the equivalence of  Flavor SU(3) analyses of $B\to PP$ decays}

\author{Yu-Ji Shi}
\email{shiyuji@ecust.edu.cn}
\affiliation{School of Physics, East China University of Science and Technology, Shanghai 200237, China}
\author{Wei Wang}
\email{wei.wang@sjtu.edu.cn}
\affiliation{State Key Laboratory of Dark Matter Physics, School of Physics and Astronomy, Shanghai Jiao Tong University,  Shanghai 200240, China}
\affiliation{Southern Center for Nuclear-Science Theory (SCNT),
Institute of Modern Physics, Chinese Academy of Sciences,
Huizhou 516000, Guangdong Province, P.R. China}
\author{Ji Xu}
\email{xuji1991@sjtu.edu.cn}
\affiliation{School of Nuclear Science and Technology, Lanzhou University, Lanzhou 730000, China}

\begin{abstract}
We conduct an SU(3) analysis of $B\to PP$ decays based on reduced matrix elements (RMEs), with $P$ being a light pseudoscalar meson excluding $\eta^{(\prime)}$.  We show that a complete basis for the $B\to PP$ decays consists of ten RMEs, where the three RMEs arise  from the electroweak penguin operators $O_{7,8}$. In the Standard Model, the relevant Wilson coefficients are small and thus can be neglected. We further demonstrate the equivalence of the RME approach with the irreducible representation amplitude (IRA) and topological diagram amplitude (TDA) methods, and derive relations between the ten RME amplitudes and corresponding IRA/TDA amplitudes. These relations lay a foundation for consistent SU(3) analyses of heavy meson decays.

\end{abstract}

\maketitle

\section{Introduction}

Flavor SU(3) symmetry has emerged as a powerful tool in the analysis of heavy meson  decays~\cite{Zeppenfeld:1980ex,Savage:1989ub,Deshpande:1994ii,He:1998rq,He:2000ys,Hsiao:2015iiu,Chau:1986du,Chau:1987tk,Chau:1990ay,Gronau:1994rj,Gronau:1995hm,Cheng:2014rfa}. It provides a framework that is independent of the detailed dynamics of factorization, allowing for a more model-independent approach to understanding the decay amplitudes. There are two main ways to implement the flavor SU(3) analysis: the irreducible representation amplitude (IRA) method and the topological diagram amplitude (TDA) method. In Refs.~\cite{He:2018php,He:2018joe}, it is explicitly demonstrated that IRA and TDA methods yield equivalent physical results in the SU(3) limit, as they are based on the same underlying flavor SU(3) symmetry.
For example, in $B \to PP$ decays where $P$ represents a light pseudoscalar meson (excluding $\eta^{(\prime)}$), there are five independent amplitudes involving the CKM matrix elements~$V_{ub}V_{ud/s}^*$~and another five independent amplitudes involving~$V_{tb}V_{td/s}^*$, resulting in a total of ten amplitudes.   Relations between these amplitudes with QCD factorization approach have been established and applied in many weak decays of heavy hadrons~\cite{Huber:2021cgk,BurgosMarcos:2025xja,Wang:2025bdl,Jia:2024pyb,Sun:2024mmk,Wang:2024ztg,Han:2024kgz,Wang:2023uea,Zhou:2023lbc,Qin:2022nof,Zhong:2022exp,Xing:2022aij,Qin:2022nof,Hsiao:2021nsc,Bhattacharya:2021ndt,Han:2021gkl,Shi:2020gfp,Roy:2020nyx,Geng:2020fng,Wang:2020gmn,Jia:2019zxi,Xing:2019hjg,Zhu:2018epc,Xing:2018bqt,Wang:2018utj,Wu:2025hnh,Geng:2024sgq,Geng:2023pkr,Geng:2018upx,Geng:2018bow,Huang:2021aqu,Liu:2023zvh,Xing:2024nvg,Lu:2016ogy,Wang:2017azm,Huang:2021jxt}.

Recently in Ref.~\cite{Berthiaume:2023kmp}, an SU(3) analysis based on reduced matrix elements (RMEs) for $B\to PP$ (excluding $\eta^{(\prime)}$) decays  was conducted. The analysis  aimed to test the flavor SU(3) symmetry in the context of the standard model by performing global fits to experimental data. However, in the analysis there are only seven independent matrix elements in the standard model, three fewer than in IRA and TDA. Ref.~\cite{Bhattacharya:2025wcq} has  conducted an  analysis of $B\to PP$ decays by including the $\eta^{(\prime)}$ in the final state, where there are also only seven independent matrix elements for decays excluding a $\eta^{(\prime)}$ meson.  The difference in independent SU(3) amplitudes  may have posed a potential  discrepancy  for the SU(3) symmetry analysis of heavy meson decays that needs to be clarified.   In addition, it would be valuable to establish a connection between these realizations of SU(3) analysis.

In this short note, we aim to  make some clarifications and  bridge the gaps between the RMEs, and  IRA, TDA  approaches.  We point out that  as a complete basis  there are in total ten, {\it  not seven},   independent amplitudes  for $B\to PP$ decays excluding $\eta^{(\prime)}$ in the final state. After giving these ten RMEs,  we demonstrate the equivalence with the other two realizations of flavor SU(3) symmetry in $B\to PP$ decays, and derive the relations between the ten RME amplitudes and the corresponding ten IRA, TDA amplitudes.

The three RMEs that are neglected  in Ref.~\cite{Berthiaume:2023kmp} arise from the contributions from the  electroweak penguin operators $O_{7,8}$  as in Ref.~\cite{Berthiaume:2023kmp}. After neglecting  the $O_{7,8}$ operators, the electroweak penguins have the same structure with tree operators. Thus three RMEs can be removed through a so-called  EWP-tree relation~\cite{Neubert:1998pt,Neubert:1998jq,Gronau:1998fn}, namely the electroweak penguin amplitudes are connected to tree amplitudes through  ratios of Wilson coefficients, $(C_{9}+C_{10})/(C_1+C_2)$ and $(C_{9}-C_{10})/(C_1-C_2)$.  This relation can be made explicit   using the QCD factorization (QCDF)~\cite{Beneke:2001ev}. Since the Wilson coefficients for $O_{7,8}$ are small in the standard model, neglecting these contributions may not have sizable impact in phenomenological analysis. It should be mentioned that the contributions from $O_{7,8}$ are  generally  smaller than the typical SU(3)-breaking effects. Since this work is dedicated to demonstrating the equivalence of the RME, IRA and TDA approaches within exact flavor SU(3) symmetry,  the SU(3)-breaking effects and their impact in phenomenological analysis will not be considered in this work.

The rest of this work is organized as follows. We will briefly give the framework for IRA and TDA in Sec.~\ref{sec:IRA_TDA}. In Sec.~\ref{sec:RME_IRA_TDA}, the complete RMEs are given and relations with IRA and TDA are established through a matching of the meson matrix.  In Sec.~\ref{sec:RME_Simplification},  based on the approximation to neglect the $O_{7,8}$ we give some explanations to remove the three RMEs. Some summaries are collected in the last section.

\section{IRA and TDA}
\label{sec:IRA_TDA}

Before  discussing the relation with RMEs, we will briefly give the framework for IRA and TDA.
In our previous studies~\cite{He:2018php,He:2018joe}, the IRA and TDA amplitudes for $B\to PP$ decays ($P$ representing  a light pseudoscalar meson excluding $\eta^{(\prime)}$) are constructed and their equivalence is established.

Charmless $B \to PP$ decays induced by the transitions ${b} \to q $, $q = d, s$ are governed by the weak Hamiltonian~\cite{Buchalla:1995vs}
\begin{eqnarray}
H_W &=& \frac{G_F}{\sqrt{2}} \sum_{q=d,s}\left(\lambda_u^{(q)}\left[C_1O_1 +C_2O_2\right] - \lambda_t^{(q)} \sum_{i=3}^{10} C_i O_i  \right), \label{eq:HW}
\end{eqnarray}
where $\lambda_{u}^{(q)} = V_{ub} V_{uq}^*$ and $\lambda_{t}^{(q)} = V_{tb} V_{tq}^*$ with $q=d,s$. Here the $C_i$ ($i = 1$-10) are Wilson coefficients.  $O_{i}$ is a four-quark operator or a moment type operator. The four-quark operators $O_i$ are given as follows:
\begin{eqnarray}
 O_1 =(\bar u^j b^i)_{V-A}  (\bar q^i u^j)_{V-A} , && O_2 =(\bar u b)_{V-A} (\bar q u)_{V-A} , \nonumber\\
 O_3= (\bar q b)_{V-A} \sum_{q'} (\bar q'q')_{V-A}, && O_4= (\bar q^i b^j)_{V-A} \sum_{q'} (\bar q^{\prime j}q^{\prime i})_{V-A}, \nonumber\\
 O_5= (\bar q b)_{V-A} \sum_{q'} (\bar q'q')_{V+A}, && O_6= (\bar q^i b^j)_{V-A} \sum_{q'} (\bar q^{\prime j}q^{\prime i})_{V+A}, \nonumber\\
 O_7=\frac{3}{2} (\bar q b)_{V-A} \sum_{q'} e_{q'}(\bar q'q')_{V+A}, && O_8= \frac{3}{2}(\bar q^i b^j)_{V-A} \sum_{q'}e_{q'} (\bar q^{\prime j}q^{\prime i})_{V+A}, \nonumber\\
 O_9=\frac{3}{2} (\bar q b)_{V-A} \sum_{q'}e_{q'}  (\bar q'q')_{V-A}, &&  O_{10}= \frac{3}{2}(\bar q^i b^j)_{V-A} \sum_{q'}e_{q'}  (\bar q^{\prime j}q^{\prime i})_{V-A}. \label{operators}
\end{eqnarray}
The electroweak penguin operators can be re-expressed as:
\begin{eqnarray}
(\bar q b)_{V-A} \sum_{q'} e_{q'}(\bar q'q')_{V\pm A}=  (\bar q b)_{V-A} (\bar uu)_{V\pm A} - \frac{1}{3} (\bar q b)_{V-A} \sum_{q'} (\bar q'q')_{V\pm A}, \label{relation}
\end{eqnarray}
where the second part can be incorporated into the penguins transforming as a triplet under SU(3).  The contributions from $\bar q b \bar uu $ has the similar structure with   tree operators, and for convenience we define the new operators:
\begin{eqnarray}
 O_7'=\frac{3}{2} (\bar q b)_{V-A} (\bar uu)_{V+A}, \;\;\;\; && O_8'= \frac{3}{2}(\bar q^i b^j)_{V-A}   (\bar u^{j}u^{i})_{V+A}, \nonumber\\
 O_9'=\frac{3}{2} (\bar q b)_{V-A}    (\bar uu)_{V-A} = \frac{3}{2}O_1, \;\;\;\; &&  O_{10}'= \frac{3}{2}(\bar q^i b^j)_{V-A}   (\bar u^{j}u^{i})_{V-A}= \frac{3}{2}O_2. \label{new_operators}
\end{eqnarray}
We will name the contributions from these operators, instead the original contributions from $O_{7-10}$, as the electroweak penguins.

The decay amplitudes of $B\to PP$ decays in IRA and TDA can be written as
\begin{eqnarray}
&&{\cal A}^{IRA} =\lambda_{u}^{(q)} {\cal A}^{IRA}_t  - \lambda_{t}^{(q)} {\cal A}^{IRA}_p\;,\nonumber\\
&&{\cal A}^{TDA} = \lambda_{u}^{(q)} {\cal A}^{TDA}_t  - \lambda_{t}^{(q)} {\cal A}^{TDA}_p\;.
\end{eqnarray}
In SU(3) representation, the $B$ meson and the octet meson are described by following irreducible tensors:
\begin{eqnarray}
 B_i&=&(B(\bar b u), B(\bar b d), B(\bar b s)),\nonumber\\ \nonumber\\
 (M_8)^i_j&=&\begin{pmatrix}
 \frac{\pi^0}{\sqrt{2}}+\frac{\eta_8}{\sqrt{6}}  &\pi^+ & K^+\\
 \pi^-&-\frac{\pi^0}{\sqrt{2}}+\frac{\eta_8}{\sqrt{6}}&{K^0}\\
 K^-&\overline K^0 &-2\frac{\eta_8}{\sqrt{6}} ,\label{eq:repBM}
 \end{pmatrix}.
\end{eqnarray}
The effective Hamiltonian in Eq.(\ref{eq:HW}) should be decomposed to a vector $H_{\bf \bar 3}$, a traceless
tensor with antisymmetric upper indices: $H_{\bf6}$, and a
traceless tensor with symmetric   upper indices: $H_{\bf{\overline{15}}}$.
For the $\Delta S=0 (b\to d)$ decays, the non-zero components of the effective Hamiltonian are~\cite{Savage:1989ub,He:2000ys,Hsiao:2015iiu}:
\begin{eqnarray}
 (H_{\bf \bar3})^2=1,\;\;\;(H_{6})^{12}_1=-(H_{6})^{21}_1=(H_{6})^{23}_3=-(H_{6})^{32}_3=1,\nonumber\\
 2(H_{\overline{15}})^{12}_1= 2(H_{\overline{15}})^{21}_1=-3(H_{\overline{15}})^{22}_2=
 -6(H_{\overline{15}})^{23}_3=-6(H_{\overline{15}})^{32}_3=6,\label{eq:H3615_bb}
\end{eqnarray}
and  all other remaining components are zero. For the $\Delta S=-1(b\to s)$
decays the nonzero components of the $H_{\bf{\bar 3}}$, $H_{\bf 6}$,
$H_{\bf{\overline{15}}}$ can be  obtained from Eq.~\eqref{eq:H3615_bb}
by the exchange  $2\leftrightarrow 3$ corresponding to $d \leftrightarrow s$.

To obtain IRA amplitudes for $B \to PP$  decays, one takes the tensors in Eq.~\eqref{eq:repBM} and Eq.~\eqref{eq:H3615_bb}, and uses them to construct all the possible SU(3) invariant terms:
\begin{eqnarray}\label{nonisodecomp}
 {\cal A}^{IRA}_t &=&A_3^T B_i (H_{\bar 3})^i (M_8)_k^j(M_8)_j^k +C_3^T B_i (M_8)^i_j (M_8)^j_k (H_{\bar 3})^k \nonumber\\
  &&\;\;\;
  +A_6^T B_i (H_{ 6})^{ij}_k (M_8)_j^l(M_8)_l^k
  +C_6^T B_i (M_8)^i_j (H_{ 6})^{jl}_k (M_8)_l^k  \nonumber\\
  && \;\;\;
  +A_{15}^T B_i (H_{\overline{15}})^{ij}_k (M_8)_j^l(M_8)_l^k
  +C_{15}^T B_i (M_8)^i_j (H_{\overline{15}})^{jk}_l (M_8)_k^l . \label{eq:IRAamps}
\end{eqnarray}
The penguin amplitudes $A^{IRA}_p$ can be obtained by the replacements  $A_i^T\to A_i^P$,  and   $C_i^T\to C_i^P$.
Expanding the above ${\cal A}_t^{IRA}$, one can obtain the $B\to PP$ amplitudes. Note that the amplitude $A_{6}^T$ can be absorbed into   $C_6^T$ with the following redefinition:
\begin{eqnarray}
C_{6}^{T\prime}= C_{6}^T-A_{6}^T.
\end{eqnarray}

In TDA, the effective Hamiltonian is represented by $H^{ij}_k$. The non-zero components are $\bar H^{12}_1= 1$ for $\Delta S=0$ and $\bar H^{13}_1=1$ for $\Delta S=-1$.
The tree amplitude of $B\to PP$ in TDA is given as
\begin{eqnarray}
{\cal A}^{TDA}_{t} &=&  T~  B_i \bar H^{jl}_k (M_8)^{i}_j    (M_8)^k_l   +C~  B_i \bar H^{lj}_k (M_8)^{i}_j  (M_8)^k_l + A~ B_i \bar H^{il}_j   (M_8)^j_k (M_8)^{k}_l   \nonumber\\
  && + E~  B_i  \bar H^{li}_j (M_8)^j_k (M_8)^{k}_l +T_{P} B_i \bar H^{lk}_{l} (M_8)^{i}_j   (M_8)^j_k  + T_{PA} B_i \bar H^{li}_{l}  (M_8)^j_k (M_8)^{k}_j .\label{eq:tree_TDA}
\end{eqnarray}

Comparing the IRA and TDA amplitudes,  one can derive the relations  between them:
\begin{eqnarray}
&&A_3^T= -\frac{A}{8} + \frac{3E}{8}+T_{PA}, \;
C_3^T=  \frac{1}{8} ({3A-C-E+3T})+T_P, \;\;\; D_3^T=  \frac{1}{8} (3C-T),\nonumber\\
&& C_6^{\prime T}=  \frac{1}{4}(-A-C+E+T), \;\;\;
A_{15}^T=  \frac{A+E}{8},
C_{15}^T=  \frac{C+T}{8}. \label{eq:relation_TDA2IRA}
\end{eqnarray}
Similar relations exist for penguin contributions with $V_{tb}V_{tq}^*$.

\section{Reduced Matrix Elements, and the equivalence with  IRA and TDA}
\label{sec:RME_IRA_TDA}

 In contrast to the IRA and TDA analysis, a notable aspect based on the  RMEs highlighted in the study by Ref.~\cite{Berthiaume:2023kmp} is the use of SU(3) decomposition  of  the two final-state meson octets: they couple through the symmetric $({\bf 8} \times {\bf 8})_s = {\bf 1} + {\bf 8} + {\bf 27}$ channel. The complete RMEs for $B\to PP$ decays should  be constructed as:
\begin{eqnarray}
\lambda_u^{(q)} &:& A_1^u = \langle {\bf 1} || {\bf\overline{3}} || {\bf 3} \rangle ~,~
A_8^u  = \langle {\bf 8} || \bf\overline{3} || {\bf 3} \rangle ~,~ \nonumber \\
& & R_8^u = \langle {\bf 8} || {\bf 6} || {\bf 3} \rangle ~,~
P_8^u = \langle {\bf 8} || \overline{\bf 15} || {\bf 3} \rangle ~,~   P_{27}^u  = \langle {\bf 27} || \overline{\bf 15} || {\bf 3} \rangle~. \nonumber \\
\lambda_t^{(q)} &:& A_1^t = \langle {\bf 1} || {\bf\overline{3}} || {\bf 3} \rangle ~,~
A_8^t = \langle {\bf 8} || {\bf\overline{3}} || {\bf 3} \rangle ~, \nonumber \\
  & & R_8^t = \langle {\bf 8} || {\bf 6} || {\bf 3} \rangle ~,~
P_8^t = \langle {\bf 8} || \overline{\bf 15} || {\bf 3} \rangle ~,~   P_{27}^t = \langle {\bf 27} || \overline{\bf 15} || {\bf 3} \rangle~, \label{eq:RMEampsCorrect}
\end{eqnarray}
where we have added the superscript $u,t$ corresponding to the CKM factors $V_{ub}V_{ud/s}^*,  V_{tb}V_{td/s}^*$.  For the amplitudes involving the CKM elements $V_{ub}V_{ud/s}^*$, there are five independent amplitudes due to the SU(3) decomposition of the $O_{1,2}$ operators. For the amplitudes involving~$V_{tb}V_{td/s}^*$, since the operators $O_{7-10}$ behave similarly with $O_{1,2}$ under the SU(3) flavor transformations, thus there are five amplitudes in comparison with the case of $V_{ub}V_{ud/s}^*$. Notice that the operators $O_{7,8}$ have different structures with $O_{1,2}$, thus one can not make any relation between these contributions.

In the following we will derive the relations of  these amplitudes and IRA, TDA amplitudes  in a straightforward way.     A key ingredient to realize this is the decomposition of the products of the meson octet $(M_8)_k^j(M_8)_j^k$: $(8\otimes 8)_s=1\oplus 8 \oplus 27$. Generally, it can be decomposed into as
\begin{eqnarray}
(M_8)^i_j(M_8)^k_l &=&  (MM_{27})^{ik}_{jl} + \frac{1}{5} \delta^i_l  (MM_8)^{k}_{j} + \frac{1}{5} \delta^{k}_{j}  (MM_8)^i_l + \frac{1}{8} \left(\delta^i_l \delta^{k}_{j}-\frac{1}{3}\delta^i_j \delta^{k}_{l}\right) (MM_1) + \frac{2}{5} \epsilon_{jlm}\epsilon^{ikn} (MM_8)^m_n,\label{eq:MMdecompose}
\end{eqnarray}
where $MM_1= (M_8)_k^j(M_8)_j^k$  is the singlet, the traceless octet tensor $(MM_8)_k^i$ is defined as
\begin{eqnarray}
(M_8)^i_j(M_8)^j_k &=&  (MM_8)^i_k + \frac{1}{3} \delta^i_k  (MM_1),
\end{eqnarray}
and the 27-state is denoted by the traceless and totally symmetrized tensor: $(MM_{27})^{ik}_{jl}$.

Using Eq.(\ref{eq:MMdecompose}), one can express the IRA amplitudes given in Eq.(\ref{eq:IRAamps}) by the $MM_1, MM_8, MM_{27}$ states. We use the amplitudes with the CKM $V_{ub}V_{uq}^*$ as an example and the IRA amplitudes induced by $H_{\bar 3}$ can be expressed as
\begin{eqnarray}\label{nonisodecomp}
&& A_3^T B_i (H_{\bar 3})^i (M_8)_k^j(M_8)_j^k +C_3^T B_i (M_8)^i_j (M_8)^j_k (H_{\bar 3})^k \nonumber\\
&=& A_3^T B_i (H_{\bar 3})^i (MM_1) +C_3^T B_i (MM_8)^i_k (H_{\bar 3})^k + \frac{1}{3} C_3^T B_i (MM_1) \delta^i_k (H_{\bar 3})^k \nonumber\\
&=&\left(A_3^T + \frac{1}{3} C_3^T\right) B_i (H_{\bar 3})^i (MM_1) +C_3^T B_i (MM_8)^i_k (H_{\bar 3})^k.
\end{eqnarray}
Compared with the RME amplitudes given in Eq.(\ref{eq:RMEampsCorrect}), we can derive the relations:
\begin{eqnarray}
  A_1^u = A_3^T + \frac{1}{3}C_3^T, \;\;\;\;     A_8^u = C_3^T. \label{eq:RMEandIRA1}
\end{eqnarray}
The amplitudes induced by $H_{6}$  can be expressed as
\begin{align}
 & A_{6}^{T}B_{i}(H_{6})_{k}^{ij}(M_{8})_{j}^{l}(M_8)_{l}^{k}+C_{6}^{T}B_{i}(M_{8})_{j}^{i}(H_{6})_{k}^{jl}(M_{8})_{l}^{k}\nonumber \\
= & A_{6}^{T}B_{i}(H_{6})_{k}^{ij}(MM_{8})_{j}^{k}+\frac{1}{5}C_{6}^{T}B_{i}(H_{6})_{k}^{ji}(MM_{8})_{j}^{k}+\frac{2}{5}C_{6}^{T}B_{i}(H_{6})_{k}^{jl}\epsilon_{jlm}\epsilon^{ikn}(MM_{8})_{n}^{m},
\end{align}
where some terms in Eq.(\ref{eq:MMdecompose}) do not contribute since $H_6$ is traceless and its upper indexes are anti-symmetric.
Further, we use the identity:
\begin{equation}
\epsilon_{jlm}\epsilon^{ikn}={\rm det}\begin{vmatrix}\delta_{j}^{i} & \delta_{l}^{i} & \delta_{m}^{i}\\
\delta_{j}^{k} & \delta_{l}^{k} & \delta_{m}^{k}\\
\delta_{j}^{n} & \delta_{l}^{n} & \delta_{m}^{n}
\end{vmatrix}=\delta_{j}^{i}\delta_{l}^{k}\delta_{m}^{n}+\delta_{l}^{i}\delta_{m}^{k}\delta_{j}^{n}+\delta_{m}^{i}\delta_{j}^{k}\delta_{l}^{n}-\delta_{m}^{i}\delta_{l}^{k}\delta_{j}^{n}-\delta_{j}^{i}\delta_{m}^{k}\delta_{l}^{n}-\delta_{l}^{i}\delta_{j}^{k}\delta_{m}^{n}\label{eq:simplifyEps}
\end{equation}
to simplify the last term above, and then obtain
\begin{align}
 & A_{6}^{T}B_{i}(H_{6})_{k}^{ij}(M_{8})_{j}^{l}(M_8)_{l}^{k}+C_{6}^{T}B_{i}(M_{8})_{j}^{i}(H_{6})_{k}^{jl}(M_{8})_{l}^{k}\nonumber \\
= & \left(A_{6}^{T}-C_{6}^{T}\right)B_{i}(H_{6})_{k}^{ij}(MM_{8})_{j}^{k}.
\end{align}
It can be found that only the octet final state can be produced through $H_6$, which is consistent with Eq.(\ref{eq:RMEampsCorrect}). We can conclude
\begin{eqnarray}
R_8^u=A_{6}^{T}-C_{6}^{T}. \label{eq:RMEandIRA2}
\end{eqnarray}
This demonstrates that $A_{6}^{T}$  and $C_{6}^{T}$ are not independent, a consequence of the symmetry present in the final states.  Another rationale for the dependence of  $A_{6}^{T}$  and $C_{6}^{T}$ is given in Ref.~\cite{Wang:2020gmn}.

Following a similar procedure, and using Eq.(\ref{eq:simplifyEps}), one can derive the amplitudes induced by $H_{\bar{15}}$ as:
\begin{align}
 & A_{15}^{T}B_{i}(H_{\overline{15}})_{k}^{ij}(M_{8})_{j}^{l}(M_{8})_{l}^{k}+C_{15}^{T}B_{i}(M_{8})_{j}^{i}(H_{\overline{15}})_{l}^{jk}(M_{8})_{k}^{l}\nonumber \\
= & C_{15}^{T}B_{i}(H_{\overline{15}})_{k}^{jl}(MM_{27})_{jl}^{ik}+(A_{15}^{T}+\frac{1}{5}C_{15}^{T})B_{i}(H_{\overline{15}})_{k}^{ij}(MM_{8})_{j}^{k},
\end{align}
which gives
\begin{eqnarray}
P_8^u =  A_{15}^T+\frac{1}{5}C_{15}^T,\;\;\; P_{27}^u = C_{15}^T .\label{eq:RMEandIRA3}
\end{eqnarray}

To obtain the relation between the RMEs and TDA amplitudes, one can similarly obtain the results by substituting the meson decomposition in Eq.~\eqref{eq:MMdecompose} into Eq.~\eqref{eq:tree_TDA}. However, another direct way is to make use of  the relation between IRA and TDA. Finally, we can obtain the relation between the RME and TDA amplitudes:
\begin{align}
A_{1}^{u} & =\frac{1}{24}\left(24T_{PA}-C+8 {E}+8T_{P}+3T\right)\nonumber \\
A_{8}^{u} & =\frac{1}{8}(3A-C- {E}+3T +8T_{P})\nonumber \\
R_{8}^{u} & =\frac{1}{4}(A+C- {E}-T)\nonumber \\
P_{8}^{u} & =\frac{1}{40}(5A+C+5 {E}+T)\nonumber \\
P_{27}^{u} & =\frac{1}{8}(C+T).\label{eq:REMandTDA}
\end{align}
It should be noted that the $E$ can be incorporated into the others.  Some of these relations have been given previously for instance in  Ref.\cite{Gronau:1994rj} and our results are consistent.   By definition, our TDA amplitudes $T, A, E, C$ differ by an overall minus sign compared to those given in these literatures, while the rest two TDA amplitudes are related as: $T_P=-P_{uc}, T_{PA}=-(1/2)PA_{uc}$. Substituting the relation between these two sets of TDA amplitudes into Eq.~(\ref{eq:REMandTDA}), and ignoring the overall constant factor in the definition of RME amplitudes, we can obtain results consistent with that from Refs.\cite{Gronau:1994rj}.

There are similar relations for the five penguin amplitudes. The penguin amplitudes ${\cal A}^{IRA}_p$ in IRA and ${\cal A}^{TDA}_p$ in TDA have the same form as the tree amplitudes given in  Eq.(\ref{eq:IRAamps}) and Eq.(\ref{eq:tree_TDA}), which can be obtained by replacement:
\begin{align}
A_{i}^{t}\to A_{i}^{p},\ \ \ C_{i}^{t}\to C_{i}^{p}
\end{align}
for IRA and
\begin{align}
T\to P_T,\ \ \ C\to P_C,\ \ \ A\to P_{TA},\ \ \ E\to P_{TE},\ \ \ T_P\to P,\ \ \ T_{PA}\to P_A
\end{align}
for TDA. The relations between penguin amplitudes in RME and IRA can be derived in almost the same way as those of tree amplitudes as given in Eq.(\ref{eq:RMEandIRA1}), Eq.(\ref{eq:RMEandIRA2}) and Eq.(\ref{eq:RMEandIRA3}). Similarly, one can obtain the relation between the penguin amplitudes in RME and IRA  as
\begin{align}
A_{1}^{t} & =\frac{1}{24}\left(24 P_A-P_C+8 P_{TE}+8 P+3 P_T\right)\nonumber \\
A_{8}^{t} & =\frac{1}{8}(3 P_{TA}-P_C- P_{TE}+3 P_T +8 P)\nonumber \\
R_{8}^{t} & =\frac{1}{4}(P_{TA}+P_C- P_{TE}-P_T)\nonumber \\
P_{8}^{t} & =\frac{1}{40}(5 P_{TA}+P_C+5 P_{TE}+P)\nonumber \\
P_{27}^{t} & =\frac{1}{8}(P_C+P_T).\label{eq:REMandTDA_Penguin}
\end{align}
These relations are new compared to the literature.

\section{Simplifications of RMEs}
\label{sec:RME_Simplification}

As an approximation,  the electroweak penguin contributions from $O_{7,8}$ may be  neglected due to the small Wilson coefficients. Under this approximation,   three RMEs can be removed.  The other two electroweak penguin operator $O_{9,10}$ have the same structures with tree operators.  After removing a SU(3) triplet contributions into the QCD penguins,  the $O'_{9,10}$ operators  have a similar structure with tree operators:
\begin{eqnarray}
 O_9'=\frac{3}{2} (\bar q b)_{V-A}    (\bar uu)_{V-A} = \frac{3}{2}O_1, \;\;\;\; &&  O_{10}'= \frac{3}{2}(\bar q^i b^j)_{V-A}   (\bar u^{j}u^{i})_{V-A}= \frac{3}{2}O_2, \label{new_operators_relations_tree}
\end{eqnarray}
where Fierz transformations have been adopted.

  Thereby in Ref.~\cite{Berthiaume:2023kmp},  seven RMEs are constructed for  charmless $B \to
PP$ decays:
\begin{eqnarray}
\lambda_u^{(q)} &:& A_1 = \langle {\bf 1} || {\bf\overline{3}} || {\bf 3} \rangle ~,~
A_8 = \langle {\bf 8} || {\bf\overline{3}} || {\bf 3} \rangle ~,~ \nonumber \\
\lambda_t^{(q)} &:& B_1 = \langle {\bf 1} || {\bf\overline{3}} || {\bf 3} \rangle ~,~
B_8 = \langle {\bf 8} || {\bf\overline{3}} || {\bf 3} \rangle ~, \nonumber \\
\lambda_u^{(q)}~{\&}~\lambda_t^{(q)} &:& R_8 = \langle {\bf 8} || {\bf 6} || {\bf 3} \rangle ~,~
P_8 = \langle {\bf 8} || \overline{\bf 15} || {\bf 3} \rangle ~,~   P_{27} = \langle {\bf 27} || \overline{\bf 15} || {\bf 3} \rangle~. \label{eq:RMEamps}
\end{eqnarray}
For the sextet and 15-plet operators, as one can see from the effective Hamiltonian in Eq. (1), contributions arise from both tree-level operators and electroweak operators, and six reduced matrix elements should be constructed. However in Eq.~(\ref{eq:RMEamps}), it is evident that there are  only three reduced matrix elements corresponding to the  sextet and 15-plet operators.

The impact of electroweak penguin contributions can be effectively mitigated by employing the EWP-tree relations. These relations express electroweak penguins as a combination of tree amplitudes and ratios of Wilson coefficients $C_{1,2,9,10}$, allowing for the representation of electroweak penguins in terms of tree contributions.  It should be noticed that as Ref.~\cite{Gronau:1998fn}, the sextet contributions only contain the product of $C_{1}-C_2$ due to the antisymmetry property of the two quark fields, while the 15-plet contributions only contain the $C_9+C_{10}$ due to the symmetric properties of the two quark fields.  This leads to
\begin{eqnarray}
\frac{R_8^t}{R_8^u} =  \frac{3}{2} \frac{C_9-C_{10}}{C_1-C_2}, \;\;\;
\frac{P_8^t}{P_8^u} = \frac{P_{27}^t}{P_{27}^u} =  \frac{3}{2} \frac{C_9+C_{10}}{C_1+C_2},
\end{eqnarray}
where the  factor $3/2$ arises from the renamed operator.
These ratios are scale-invariant, and thus constants, since the anomalous dimension matrices for the operators  are the same~\cite{Buchalla:1995vs}.

The above relations can be  illustrated through an explicit analysis in QCD factorization(QCDF) where some recent analyses are given in Refs.~\cite{Huber:2021cgk,BurgosMarcos:2025xja}. Based on the SU(3) analyses there are conjuncted studies in QCDF, where some dynamical effects can be extracted when compared to the experimental data.  As given by Ref.~\cite{Gronau:1994rj}, the QCDF amplitudes were connected with the TDA amplitudes, where the linear transformation relationship between them are provided.
In QCDF, the topological amplitudes for the tree operators can be given as:
\begin{eqnarray}
 T= A_{M_1M_2}^{B_q}\tilde \alpha_1, \;\;\;\;
 C= A_{M_1M_2}^{B_q}\tilde \alpha_2,\;\;\;\;
 A= B_{M_1M_2}^{B_q}\tilde b_2,\;\;\;\;
 E= B_{M_1M_2}^{B_q}\tilde b_1.
\end{eqnarray}
Similarly for the electroweak penguins, they are given as:
\begin{eqnarray}
 P_T= \frac{3}{2} A_{M_1M_2}^{B_q}\alpha_{4,EW},\;\;\;\;
 P_C= \frac{3}{2}A_{M_1M_2}^{B_q}\alpha_{3,EW},\;\;\;\;
 P_{TA}=\frac{3}{2} B_{M_1M_2}^{B_q}b_{3,EW},\;\;\;\;
 P_{TE}=\frac{3}{2} A_{M_1M_2}^{B_q}b_{4,EW},
\end{eqnarray}
where $A_{M_1M_2}^{B_q}$ and $A_{M_1M_2}^{B_q}$ are products of decay constants and transition form factors. $\tilde \alpha_i$ and  $\tilde b_i$  contain the relevant Wilson coefficients, the coefficients $\alpha_{i,EW}$ and $b_{i,EW}$ can be obtained with the replacements $C_1\to 3/2 C_9$ and $C_{2}\to 3/2 C_{10}$. It should be noticed that this replacement works only when the contributions from $O_{7,8}$ are neglected.
Making use of the relations in Eq.~\eqref{eq:REMandTDA}, one has:
\begin{eqnarray}
\frac{R_8^t}{R_8^u}  &=&\frac{P_{TA}+P_C- P_{TE}-P_T}{A+C- {E}-T}= \frac{3}{2} \frac{C_9-C_{10}}{C_1-C_2}, \nonumber\\
\frac{P_8^t}{P_8^u} &=& \frac{5P_{TA}+P_C+5 P_{TE}+P_T}{5A+C+5 {E}+T}= \frac{3}{2} \frac{C_9+C_{10}}{C_1+C_2},  \nonumber\\
\frac{P_{27}^t}{P_{27}^u} &=& \frac{P_C+P_T}{C+T} = \frac{3}{2} \frac{C_9+C_{10}}{C_1+C_2}, \;\;\;
\end{eqnarray}
This agrees with the EWP-tree relation.

\section{Summary}

In this analysis, we have pointed out that   as a complete basis there are in total ten RMEs for  $B\to PP$ decays where $P$ is a light pseudoscalar meson excluding $\eta^{(\prime)}$.  The equivalence with  the other two realizations of flavor SU(3) symmetry in $B\to PP$ decays has been established. We have derived the relations between the ten RME amplitudes and the corresponding ten IRA, TDA amplitudes for both tree and penguin amplitudes that will be useful for future analysis.

The three RMEs that are neglected  in Ref.~\cite{Berthiaume:2023kmp} arise from the contributions from the electroweak penguin operators $O_{7,8}$.   When the $O_{7,8}$ operators are neglected, these three RMEs can be incorporated into other amplitudes according to the so-called EWP-tree relation~\cite{Gronau:1998fn}.   Since the Wilson coefficients for $O_{7,8}$ are small in the standard model, neglecting these contribution may not have sizable  impact in phenomenological analysis.

 \section*{Acknowledgment }
 We are grateful to Keri Vos for valuable  discussions that triggered this   analysis.  We thank Fu-Sheng Yu for useful discussions. This work is supported in part by Natural Science Foundation of China under grant No.12125503, 12305103, 12335003 and 12475098.


\end{document}